\documentclass[reprint,amsmath,amssymb,aps,pre,longbibliography]{revtex4-1}

\usepackage{graphicx}
\usepackage{dcolumn}
\usepackage{bm}
\usepackage{amsmath}
\usepackage{color}
\usepackage[utf8]{inputenc}
\usepackage{cancel}
\usepackage{easyReview}
\usepackage{url}

\newcommand{\cvT}{\frac{\partial c_\rho}{\partial T}}
\newcommand{\cvr}{\frac{\partial c_\rho}{\partial \rho}}
\newcommand{\cvrr}{\frac{\partial (c_\rho\rho)}{\partial \rho}}
\newcommand{\difGT}{\frac{\partial \Gamma T}{\partial T}}

\begin{document}


\title{Thermal conductivity of Lennard-Jones fluids from equilibrium thermodynamics}

\author{M. Hoyuelos}
\email{hoyuelos@mdp.edu.ar}
\author{A. Alés}

\affiliation{Instituto de Investigaciones F\'isicas de Mar del Plata (IFIMAR-CONICET), Departamento de F\'isica, Facultad de Ciencias Exactas y Naturales, Universidad Nacional de Mar del Plata, Funes 3350, 7600 Mar del Plata, Argentina}

\begin{abstract}
We derive a general expression for the thermal conductivity of simple fluids within a thermodynamic transport theory (TTT) in which the Onsager coefficients of an interacting system are related to those of an ideal reference system through the ratio of the determinants of the Hessians of the relevant thermodynamic potential. The transported quantity is identified with an effective energy density proportional to the rms energy fluctuations of a small fluid cell. The result expresses the thermal conductivity as the product of its dilute-gas value and a function of equilibrium thermodynamic properties, namely the specific heats at constant density and at constant $\rho T$ and the temperature derivative of the latter, with no adjustable parameters. For hard spheres, it reduces to a previously verified expression in terms of the thermodynamic factor. The theory is tested against molecular dynamics results for the Lennard-Jones (LJ) and the truncated and shifted Lennard-Jones (LJTS) fluids in the supercritical region, for densities up to $\rho^*=0.8$, sufficiently far from the freezing line, since the theory applies only to a homogeneous stable fluid. For the LJTS fluid, the mean absolute percentage error is below 8\% for all the temperatures considered. For the LJ fluid, the predictions are compared with those of two versions of the revised Enskog theory (RET) for Mie fluids, using the same equation of state: TTT gives errors similar to those of the most recent version of RET, which is based on exchange-weighted transfer lengths, and much smaller than those of its original version, although the two approaches follow very different theoretical routes.
\end{abstract}


\maketitle

\section{Introduction}

The thermal conductivity is one of the fundamental transport coefficients of fluids, and its prediction from the interaction potential is a long-standing problem of statistical mechanics. In the dilute-gas limit, the problem is solved by the Chapman-Enskog solution of the Boltzmann equation, which gives the thermal conductivity in terms of collision integrals that depend only on the pair potential \cite{chapman,Hirschfelder}. At higher densities, the situation becomes more  complex: energy is transported not only by the free flight of particles between collisions, but also by the collisional transfer of energy between interacting particles, and correlations between particles play an essential role \cite{hansen2,dorfmanbook}.

The first and most successful extension of kinetic theory to dense fluids is the theory of Enskog for hard spheres \cite{enskog,chapman}, in which the effect of density is introduced through the radial distribution function at contact and the collisional transfer of momentum and energy across the particle diameter. The application of the Enskog theory to real fluids, or to models with soft interactions, requires the identification of an effective hard-sphere diameter, which led to the modified Enskog theory \cite{hanley,dymond}. The revised Enskog theory (RET) \cite{Beijeren2} provides a consistent extension to mixtures and, recently, Jervell and Wilhelmsen \cite{jervell} developed a RET for Mie fluids (RET-Mie), in which the radial distribution function at contact is obtained from the perturbation theory underlying the SAFT-VR Mie equation of state (EOS) \cite{lafitte}, without adjustable parameters. In a more recent work, the same authors introduced transfer lengths for the residual transport of momentum and energy beyond the dilute-gas limit, obtained from the intermolecular potential through an exchange-weighted closest approach (EWCA); the resulting theory, which we call RET-EWCA, significantly improves the predictions of viscosity and thermal conductivity from the dilute gas to the dense liquid \cite{jervell2024predicting}.

A different approach to dense liquids was followed by Khrapak and Khrapak \cite{khrapak2023}, who studied a vibrational model of heat transfer, in which atoms perform solid-like oscillations about temporary equilibrium positions and energy is transferred between neighboring quasi-layers at the frequencies of the collective modes. Approximating the vibrational spectrum by acoustic longitudinal and transverse modes, the thermal conductivity is expressed in terms of the specific heat and the sound velocities, which can be obtained from thermodynamic properties given by an EOS, without adjustable parameters. For the LJ fluid, this model reproduces the simulation data within a few percent in its applicability domain, the dense liquid between $\rho\simeq 0.6\rho_{\rm f}$, where $\rho_{\rm f}$ is the freezing density, and the fluid-solid coexistence; to cover the whole gas-liquid region, the authors proposed a practical expression based on the freezing density scaling of the transport coefficients \cite{khrapak2023}. This model is also an example of a direct relation between a transport coefficient and thermodynamic properties, although restricted to the liquid-like regime; the approach presented in this paper, instead, gives a single expression valid from the dilute gas to the dense fluid.

Other routes to the thermal conductivity of dense fluids are empirical or semi-empirical correlations of simulation data \cite{bugel,galliero,chaparro} and excess-entropy scaling \cite{rosenfeld,rosenfeld3,dyre}, in which the transport coefficients, suitably reduced, are expressed as functions of the excess entropy. 
Both approaches were recently combined by Jervell and Wilhelmsen in a framework called REST \cite{jervell2026}, in which the residual-entropy scaling relation, instead of being regressed to transport data, is obtained from RET-Mie at high temperatures and extrapolated to the liquid-like region.
For a given model, the thermal conductivity can be computed with molecular dynamics (MD) simulations, either with equilibrium methods based on the Green-Kubo relations \cite{zwanzig} or with nonequilibrium methods; for the Lennard-Jones (LJ) fluid and its truncated versions, extensive simulation results are available in stable and metastable states \cite{meier1,galliero,baidakov,chaparro}.

In this paper, we follow a different approach, based on equilibrium thermodynamics. The idea that the effect of interactions on transport coefficients can be largely described by thermodynamic factors has a long history in the theory of diffusion, starting with the work of Darken on binary diffusion in alloys \cite{darken}. In recent years, it has been shown that several transport coefficients can be expressed as the product of their ideal, low-concentration value and a function of the thermodynamic state: this was done for self-diffusion \cite{marchioni,sampayo} and viscosity \cite{marchioni2} of simple fluids, and for the transport coefficients of the hard-sphere fluid \cite{hoyuelosHS}. A general framework for this kind of description was recently developed by Di Muro and Hoyuelos \cite{dimuro-hoyuelos}: the Onsager matrix of an interacting system is equal to that of an ideal reference system, defined by the same transported variables, multiplied by the ratio of the determinants of the Hessians of the relevant thermodynamic potential in the ideal and real systems. We refer to this framework as the thermodynamic transport theory (TTT).

The aim of this paper is to apply TTT to the thermal conductivity of simple fluids. The main difficulty is the identification of the transported quantity: as we discuss below, the full internal energy density is not the appropriate variable, and an effective energy density, related to the energy fluctuations of a small fluid cell, has to be introduced. The result is a simple expression for the thermal conductivity as the product of its dilute-gas value and a function of equilibrium thermodynamic properties, with no adjustable parameters. We test this result against MD simulations of the LJ fluid and of the truncated and shifted Lennard-Jones (LJTS) fluid, in the supercritical region and for densities up to $\rho^*=0.8$, and, for the LJ fluid, we compare its performance with that of RET-Mie and RET-EWCA, using the same EOS.

The paper is organized as follows. In Sec.\ \ref{s.ttt}, we summarize the thermodynamic transport theory. In Sec.\ \ref{s.calc}, we derive the expression for the thermal conductivity. The comparison with numerical results and with RET is presented in Sec.\ \ref{s.comp}, and the conclusions are given in Sec.\ \ref{s.conclusions}. The calculation of the Hessian is presented in Appendix \ref{app.hessian}.

\section{Thermodynamic transport theory}
\label{s.ttt}

In this section, we briefly review the thermodynamic approach to transport introduced in Ref.\ \cite{dimuro-hoyuelos}, which is the starting point of the present work. The basic idea is to separate, in the transport coefficients of an interacting system, a contribution of purely kinetic origin, which is approximated by the corresponding coefficient of an ideal reference system, from a contribution that depends only on equilibrium thermodynamics and can therefore be evaluated with an equation of state (EOS).

The system is divided into small cells, each of which is in local thermodynamic equilibrium and exchanges extensive quantities with its neighbors; the rest of the system acts as a large reservoir for any given cell. The thermodynamic state of a cell is specified by a set of extensive variables $\vec{X}=(X_1,\dots,X_n)$, which are the quantities that can be transferred between neighboring cells, and by the conjugate entropic intensive parameters $\vec{Y} = \partial S/\partial \vec{X}$, so that
\begin{equation}
	dS = \vec{Y}\cdot d\vec{X}.
\end{equation}
In the linear regime, the fluxes of the transported quantities are proportional to the thermodynamic forces, $\vec{F} = -\nabla \vec{Y}$,
\begin{equation}\label{e.JLF}
	\vec{J} = - L \cdot \vec{F} = L \cdot \nabla \vec{Y},
\end{equation}
where $L$ is the (symmetric, positive definite) Onsager matrix \cite{lebon}. Since $\vec{Y}$ is a function of $\vec{X}$, the gradients of the intensive parameters are related to those of the extensive variables through the entropy Hessian,
\begin{equation}\label{e.H}
	H = \frac{\partial^2 S}{\partial \vec{X}^2} = \frac{\partial \vec{Y}}{\partial \vec{X}}, \qquad \nabla\vec{Y} = H\, \nabla \vec{X},
\end{equation}
so that Eq.\ \eqref{e.JLF} takes the form of a generalized Fick law, $\vec{J} = - D\, \nabla\vec{X}$, with diffusion matrix $D = -L H$. The Hessian $H$, which is negative definite in a thermodynamically stable state, also determines the amplitude of the equilibrium fluctuations of $\vec{X}$ in a cell, whose covariance matrix is $-H^{-1}$ (in units with $k_B=1$) \cite{callen,reichl}.

The relevant question is how $L$ is modified by the interactions. To answer it, the interacting system is compared with an ideal reference system defined by the same set of extensive variables in each cell, $\vec{X}_{\rm id} = \vec{X}$, but without interactions, with Hessian $H_{\rm id}$ and Onsager matrix $L_{\rm id}$. It was shown in Ref.\ \cite{dimuro-hoyuelos} that the two Onsager matrices are related by the ratio of the determinants of the Hessians,
\begin{equation}\label{e.L}
	L = \frac{\det H_{\rm id}}{\det H}\, L_{\rm id}.
\end{equation}
That is, the effect of the interactions on the Onsager matrix reduces to a scalar factor that depends only on equilibrium properties, while all the dynamical information is contained in $L_{\rm id}$.
Consequently, the diffusion matrix of the interacting system is
\begin{equation}
	D = \frac{\det H_{\rm id}}{\det H}\, L_{\rm id} H .
\end{equation}

Two remarks are important for the application of Eq.\ \eqref{e.L}. First, the set of variables $\vec{X}$ must contain only quantities that are actually transported by diffusive mechanisms. If an extensive variable $X'$, with conjugate parameter $Y'$, is not transported (for example, because of an external constraint or because its evolution is governed by a different mechanism), the relevant thermodynamic potential is no longer the entropy, but the Massieu function obtained by a Legendre transform, $\Phi = S - X' Y'$. In that case, $H$ in Eq.\ \eqref{e.L} is the Hessian of $\Phi$. Second, the transported variables of the ideal system must be identical, as functions of the state, to those of the real system; this requirement plays a central role in the derivation presented in the next section.

Equation \eqref{e.L} reproduces well-known results for binary diffusion in solids and for collective diffusion on surfaces \cite{dimuro-hoyuelos}. In the same spirit of expressing a transport coefficient as the product of its ideal value and a thermodynamic function, related expressions were obtained for self-diffusion \cite{marchioni,sampayo} and viscosity \cite{marchioni2} of simple fluids, and the method was applied to the transport coefficients of the hard-sphere (HS) fluid in Ref.\ \cite{hoyuelosHS}.

The purpose of this paper is to use the framework summarized above to derive a general expression for the thermal conductivity of a simple fluid in terms of its ideal-gas value and thermodynamic quantities, and to test it against molecular dynamics results for the Lennard-Jones (LJ) and the truncated and shifted Lennard-Jones (LJTS) fluids.

\section{Calculation of thermal conductivity}
\label{s.calc}

\subsection{Thermodynamic variables}

We consider a one-component simple fluid in continuous space. The starting point is the Gibbs relation written in terms of densities, which, for a small cell of fixed volume $V$, reads
\begin{equation}\label{e.ds}
	ds = \frac{1}{T}\, du - \frac{\mu}{T}\, d\rho,
\end{equation}
where $s=S/V$, $u=U/V$, and $\rho=N/V$ are, respectively, the entropy, internal-energy, and particle-number densities, $T$ is the temperature, and $\mu$ is the chemical potential per particle. We use units with $k_B=1$. The usual local-equilibrium hypothesis is assumed: the cell is small compared with the characteristic length scale of the perturbations and, at the same time, large enough to contain many particles, so that Eq.\ \eqref{e.ds} holds locally.

Heat conduction is a transport process of internal energy driven by a temperature gradient. However, the internal energy density $u$ is not the appropriate variable to be used directly in Eq.\ \eqref{e.L}. The reason is that only a small fraction of the internal energy of a cell participates in the exchange processes with its neighbors, and the dependence on the thermodynamic state of this fraction is not the same as that of $u$ itself. Since Eq.\ \eqref{e.L} involves the Hessian of the thermodynamic potential, i.e., the second derivatives with respect to the transported variables, this difference is not irrelevant. We therefore introduce an effective energy density, $x$, which quantifies the portion of internal energy, per unit volume, that takes part in the transport process. It is defined in such a way that its flux coincides with the flux of internal energy, $\vec{J}_x = \vec{J}_u$, while its magnitude, which is associated with fluctuations, is much smaller than $u$.

We propose that the characteristic scale of the internal energy that is exchanged between neighboring cells is given by the root-mean-square (rms) fluctuation of the internal energy of a cell. In the canonical ensemble, $\langle (\Delta U)^2 \rangle = T^2 N c_\rho$, where $c_\rho$ is the specific heat per particle at constant volume (or, equivalently for fixed $N$, at constant density). The corresponding rms fluctuation of the energy density is $u_{\rm rms}=\rho T \sqrt{c_\rho/N}$. Then,
\begin{equation}\label{e.x}
	x = \epsilon' u_{\rm rms} = \epsilon\, \rho T \sqrt{\frac{c_\rho}{c_\rho^{\rm id}}},
\end{equation}
where $\epsilon'$ is a proportionality constant and the constant factor $\sqrt{c_\rho^{\rm id}/N}$, with $N$ the (fixed) number of particles of reference in a cell, was absorbed into $\epsilon = \epsilon'\sqrt{c_\rho^{\rm id}/N}$. The parameter $\epsilon$ is independent of the thermodynamic state; as we show below, it cancels out in the final result, so that its precise value is not needed. The numerical factor was chosen so that, in the ideal gas, where $c_\rho = c_\rho^{\rm id}$, we simply have $x = \epsilon \rho T$.

\subsection{Thermodynamic potential and Onsager coefficient}

As mentioned in Sec.\ \ref{s.ttt}, the procedure developed in Ref.\ \cite{dimuro-hoyuelos} to obtain Eq.\ \eqref{e.L} requires the transported quantities to be the same functions of the state variables in the real and in the ideal systems. This is not the case for $x$, since Eq.\ \eqref{e.x} depends on $c_\rho$, which differs from its ideal value. For this reason, we first work with the auxiliary variable $\tilde{x}$, defined as the expression that $x$ takes in the ideal limit,
\begin{equation}\label{e.xt}
	\tilde{x} = \epsilon \rho T,
\end{equation}
which has the same functional form in both systems. The connection between $\tilde{x}$ and $x$ is established in Sec.\ \ref{s.conductivity}.

To write the Gibbs relation in terms of $\tilde{x}$ we use the differentials
\begin{align}
	d\tilde{x} &= \epsilon \rho \,dT + \epsilon T\, d\rho, \label{e.dxt} \\
	du &= \rho c_\rho\, dT + (\mu - T \mu_T)\,  d\rho, \label{e.durT}
\end{align}
where $\mu_T = \left(\partial \mu/\partial T\right)_\rho$. In Eq.\ \eqref{e.durT}, we have used that $(\partial u/\partial T)_\rho = \rho c_\rho$ and that $(\partial u/\partial \rho)_T = \mu - T \mu_T$; the latter follows from $u = f + Ts$, where $f$ is the Helmholtz free energy density, together with $(\partial f/\partial\rho)_T = \mu$ and the Maxwell relation $(\partial s/\partial\rho)_T = -\mu_T$. Eliminating $dT$ between Eqs.\ \eqref{e.dxt} and \eqref{e.durT}, we get
\begin{equation}
	du = \frac{c_\rho}{\epsilon}\, d\tilde{x} + (\mu - T\mu_T - Tc_\rho)\, d\rho.
\end{equation}
Replacing in Eq.\ \eqref{e.ds}, the differential of the entropy density in terms of $\tilde{x}$ and $\rho$ is
\begin{equation}\label{e.dsxt}
	ds = \frac{c_\rho}{\epsilon T}\, d\tilde{x} - (\mu_T+c_\rho)\, d\rho.
\end{equation}
The intensive parameter conjugate to $\tilde{x}$ is, therefore,
\begin{equation}\label{e.Yt}
	\tilde{Y} = \left(\frac{\partial s}{\partial \tilde{x}}\right)_\rho = \frac{c_\rho}{\epsilon T}.
\end{equation}

The particle density does not take part in the transport process under consideration: its evolution is not diffusive (in a one-component fluid there is no particle diffusion relative to the center of mass), but it is determined by the continuity equation together with the hydrodynamic velocity field. Following the prescription of Sec.\ \ref{s.ttt}, we eliminate $\rho$ as a transported variable through the Legendre transform
\begin{equation}
	\phi = s + \rho (\mu_T+c_\rho),
\end{equation}
whose differential is
\begin{equation}\label{e.dphi}
	d\phi = \frac{c_\rho}{\epsilon T}\, d\tilde{x} + \rho\, d(\mu_T+c_\rho).
\end{equation}
In this formulation there is only one transported quantity, $\tilde{x}$, and the Onsager matrix reduces to a scalar, $L_{\tilde{x}}$. Other transport coefficients, such as the viscosity or the self-diffusion coefficient, require different choices of transported variables; see Ref.\ \cite{hoyuelosHS}.

The Hessian of $\phi(\tilde{x},\mu_T+c_\rho)$ and its determinant are calculated in Appendix \ref{app.hessian}. Using the ratio of the determinants in the real and ideal cases in Eq.\ \eqref{e.L}, we obtain
\begin{equation}\label{e.LLid}
	L_{\tilde{x}} = \frac{c_\rho^{\rm id} \left[c_{\rho T} - T \left(\frac{\partial c_{\rho T}}{\partial T}\right)_{\rho T}\right]}{c_{\rho T}^{\rm id} \left(c_\rho - T\cvT\right)} L_{\tilde{x}}^{\rm id},
\end{equation}
where
\begin{equation}\label{e.crhoT}
	c_{\rho T} = c_\rho +  \frac{\partial (ZT)}{\partial T}
\end{equation}
is the specific heat per particle at constant $\rho T$, $Z = P/(\rho T)$ is the compressibility factor, and $c_{\rho T}^{\rm id} = c_\rho^{\rm id} + 1 = 5/2$ (see Appendix \ref{app.hessian}). Here and in the following, partial derivatives with respect to $T$ and $\rho$ without subscripts are taken at constant $\rho$ and constant $T$, respectively. In an ideal gas, $\rho T$ is proportional to the pressure, so that $c_{\rho T}^{\rm id}$ coincides with the specific heat at constant pressure; for an interacting fluid, $c_{\rho T}$ and the specific heat at constant pressure are different quantities.

The result \eqref{e.LLid} can also be written in terms of the curvature of the entropy per particle, $s' = s/\rho$, as a function of temperature. Since $c_\rho = T (\partial s'/\partial T)_\rho$ and $c_{\rho T} = T (\partial s'/\partial T)_{\rho T}$, we have $c_\rho - T\, \partial c_\rho/\partial T = - T^2 (\partial^2 s'/\partial T^2)_\rho$ and an analogous relation at constant $\rho T$; then,
\begin{equation}\label{e.LS}
	L_{\tilde{x}} = \frac{c_\rho^{\rm id} \left(\frac{\partial^2 s'}{\partial T^2}\right)_{\rho T}}{c_{\rho T}^{\rm id}\left(\frac{\partial^2 s'}{\partial T^2}\right)_\rho} L_{\tilde{x}}^{\rm id}.
\end{equation}
This form makes explicit that the thermodynamic factor modifying the Onsager coefficient is given by the ratio of the curvatures of the entropy along two different paths in the $(\rho,T)$ plane.

\subsection{Diffusivity of the effective energy}

The flux of $\tilde{x}$ is obtained from Eq.\ \eqref{e.JLF}. For a gradient of $\tilde{x}$ at constant density,
\begin{equation}
	\vec{J}_{\tilde{x}} =  L_{\tilde{x}} \nabla \tilde{Y} =  L_{\tilde{x}} \left(\frac{\partial \tilde{Y}}{\partial \tilde{x}}\right)_\rho \nabla \tilde{x} = - D_{\tilde{x}} \nabla \tilde{x},
\end{equation}
where $\tilde{Y}$ is given by Eq.\ \eqref{e.Yt} and $D_{\tilde{x}}$ is the diffusivity of $\tilde{x}$.
For the ideal system, $\tilde{Y}_{\rm id} = c_\rho^{\rm id}/(\epsilon T)$ and, by definition, $\tilde{x}_{\rm id} = \tilde{x}$; then,
\begin{equation}
	\vec{J}_{\tilde{x}}^{\rm id} =  L_{\tilde{x}}^{\rm id} \nabla \tilde{Y}_{\rm id} =  L_{\tilde{x}}^{\rm id} \left(\frac{\partial \tilde{Y}_{\rm id}}{\partial \tilde{x}}\right)_\rho \nabla \tilde{x} = - D_{\tilde{x}}^{\rm id} \nabla \tilde{x}.
\end{equation}
Using Eqs.\ \eqref{e.xt} and \eqref{e.Yt}, the derivatives of the intensive parameters are
\begin{align}
	\left(\frac{\partial \tilde{Y}}{\partial \tilde{x}}\right)_\rho &= \frac{1}{\epsilon^2\rho}\frac{\partial (c_\rho/T)}{\partial T} = - \frac{c_\rho - T \cvT}{\epsilon^2\rho T^2}, \\
	\left(\frac{\partial \tilde{Y}_{\rm id}}{\partial \tilde{x}}\right)_\rho &= -\frac{c_\rho^{\rm id}}{\epsilon^2\rho T^2},
\end{align}
so that the ratio of the diffusivities is
\begin{equation}\label{e.Dxtilde}
	\frac{D_{\tilde{x}}}{D_{\tilde{x}}^{\rm id}} = \frac{L_{\tilde{x}}}{L_{\tilde{x}}^{\rm id}} \frac{1}{c_\rho^{\rm id}}\left(c_\rho - T \cvT\right).
\end{equation}
Note that the factor $c_\rho - T\, \partial c_\rho/\partial T$, which is positive in a stable state because it is proportional to $-(\partial^2 s'/\partial T^2)_\rho$, cancels with the denominator of Eq.\ \eqref{e.LLid}.

\subsection{Thermal conductivity}
\label{s.conductivity}

The result \eqref{e.Dxtilde} gives information about the transport of the auxiliary variable $\tilde{x}$; we now use it to obtain the transport properties of the effective energy density $x$. From Eqs.\ \eqref{e.x} and \eqref{e.xt}, the relation between both quantities is $x = \tilde{x} R$, where, to simplify the notation, we have introduced
\begin{equation}
	R=\sqrt{\frac{c_\rho}{c_\rho^{\rm id}}}.
\end{equation}
The flux of $x$, written in terms of its own diffusivity $D_x$, is
\begin{equation}\label{e.Jx}
	\vec{J}_x = - D_x \nabla x = -D_x \left( R + \tilde{x} \frac{\partial R}{\partial \tilde{x}}\right) \nabla \tilde{x},
\end{equation}
where the derivative of $R$ is taken at constant density. On the other hand, $x$ and $\tilde{x}$ differ by the local factor $R$, which is a function of the thermodynamic state of the cell. The amount of $x$ carried by a given exchange process is $R$ times the corresponding amount of $\tilde{x}$, so that the fluxes are related by
\begin{equation}
	\vec{J}_x = R\, \vec{J}_{\tilde{x}} = - R\, D_{\tilde{x}} \nabla \tilde{x}.
\end{equation}
Comparing with Eq.\ \eqref{e.Jx}, we get the following relation between the two diffusivities:
\begin{equation}
	D_{\tilde{x}} = D_x \left( 1 + \frac{\tilde{x}}{R} \frac{\partial R}{\partial \tilde{x}}\right).
\end{equation}
In the ideal case, $R=1$ and $D_{\tilde{x}}^{\rm id} = D_x^{\rm id}$; therefore,
\begin{equation}\label{e.DDDD}
	\frac{D_x}{D_x^{\rm id}} = \frac{D_{\tilde{x}}}{D_{\tilde{x}}^{\rm id}} \left( 1 + \frac{\tilde{x}}{R} \frac{\partial R}{\partial \tilde{x}}\right)^{-1}.
\end{equation}

By definition, the flux of $x$ is equal to the flux of internal energy, i.e., to the heat flux in a fluid at rest. It can thus be written in terms of the thermal conductivity $\lambda$ and the temperature gradient through Fourier's law:
\begin{equation}\label{e.Fourier}
	\vec{J}_x = - \lambda \nabla T.
\end{equation}
Since density is not a transported variable in the present description, we consider a temperature gradient at uniform density, so that $\nabla \tilde{x} = (\partial \tilde{x}/\partial T)_\rho \nabla T = \epsilon \rho\, \nabla T$. Using this relation in Eq.\ \eqref{e.Jx} and comparing with Eq.\ \eqref{e.Fourier}, we get
\begin{equation}
	\lambda = D_x \left( R + \tilde{x} \frac{\partial R}{\partial \tilde{x}}\right) \frac{\partial \tilde{x}}{\partial T},
\end{equation}
whose value in the ideal case is
\begin{equation}
	\lambda_{\rm id} = D_x^{\rm id} \frac{\partial \tilde{x}}{\partial T}.
\end{equation}
Here, $\lambda_{\rm id}$ is the thermal conductivity of the ideal reference system at the same density and temperature, i.e., the dilute-gas value, which, as is well known, is independent of density. The ratio of thermal conductivities is then
\begin{equation}
	\frac{\lambda}{\lambda_{\rm id}} = \frac{D_x}{D_x^{\rm id}} \left( R + \tilde{x} \frac{\partial R}{\partial \tilde{x}}\right).
\end{equation}
Combining this result with Eq.\ \eqref{e.DDDD}, the terms involving the derivative of $R$ cancel out, and we obtain the simple relation
\begin{equation}
	\frac{\lambda}{\lambda_{\rm id}} = \frac{D_{\tilde{x}}}{D_{\tilde{x}}^{\rm id}}\, R.
\end{equation}
Finally, using Eqs.\ \eqref{e.Dxtilde} and \eqref{e.LLid}, we arrive at the main result of this paper:
\begin{equation}\label{e.lambda_gen}
	\frac{\lambda}{\lambda_{\rm id}} = \frac{1}{c_{\rho T}^{\rm id}} \left[c_{\rho T} - T \left(\frac{\partial c_{\rho T}}{\partial T}\right)_{\rho T}\right] \sqrt{\frac{c_\rho}{c_\rho^{\rm id}}},
\end{equation}
with $c_\rho^{\rm id} = 3/2$ and $c_{\rho T}^{\rm id} = 5/2$ for a monoatomic fluid. The parameter $\epsilon$ does not appear in Eq.\ \eqref{e.lambda_gen}: the thermal conductivity of the interacting fluid is given by the product of the dilute-gas value and a function that depends only on equilibrium thermodynamic properties, with no adjustable parameters.

For practical calculations, it is convenient to express the factor in square brackets in terms of derivatives at constant density and constant temperature. As shown in Appendix \ref{app.hessian},
\begin{equation}\label{e.bracket}
	c_{\rho T} - T \left(\frac{\partial c_{\rho T}}{\partial T}\right)_{\rho T} = c_\rho - T\cvT + 2\rho \cvr + \difGT,
\end{equation}
where $\Gamma = \partial (\rho Z)/\partial \rho$ is the thermodynamic factor. Equation \eqref{e.lambda_gen} thus requires the specific heat and its first derivatives with respect to temperature and density, together with the thermodynamic factor and its temperature derivative. When the EOS is given in terms of the Helmholtz free energy, these quantities involve derivatives of the free energy up to third order.

\subsection{Hard-sphere limit}

As a consistency check, let us consider the HS fluid. In this case, the internal energy is purely kinetic, so that $c_\rho = c_\rho^{\rm id} = 3/2$, and the compressibility factor $Z$ depends only on density; then, from Eq.\ \eqref{e.crhoT}, $c_{\rho T} = 3/2 + Z$. Using the identity
\begin{equation}\label{e.operator}
	\rho \frac{\partial}{\partial \rho} - T \frac{\partial}{\partial T} = -T \left(\frac{\partial}{\partial T}\right)_{\rho T},
\end{equation}
which follows from the fact that $d\rho/\rho = - dT/T$ at constant $\rho T$, we get $-T (\partial c_{\rho T}/\partial T)_{\rho T} = \rho\, \partial Z/\partial \rho$, and Eq.\ \eqref{e.lambda_gen} reduces to
\begin{equation}\label{e.lambda}
	\frac{\lambda}{\lambda_{\rm id}} = \frac{2}{5}\Gamma + \frac{3}{5},
\end{equation}
with $\Gamma = Z + \rho\, \partial Z/\partial\rho$. The validity of this result for the HS system was verified in Ref.\ \cite{hoyuelosHS} by comparison with molecular dynamics data. Equation \eqref{e.lambda_gen} is the generalization of Eq.\ \eqref{e.lambda} to fluids with soft interactions and attractive forces, in which the specific heat depends on the thermodynamic state and the compressibility factor depends on temperature. In the following section, we test it for the LJ and LJTS fluids.

\section{Comparison with numerical results}
\label{s.comp}

\subsection{Models and ideal-gas thermal conductivity}

We compare the theoretical prediction of Eq.\ \eqref{e.lambda_gen} with numerical results of the thermal conductivity for the LJ fluid and for the LJTS fluid with cutoff radius $r_c = 2.5 \sigma$. The LJ interaction potential as a function of the interparticle distance $r$ is
\begin{equation}
	\Phi_{\rm LJ}(r)  = 4 \varepsilon \left[ \left( \frac{\sigma}{r}\right)^{12} - \left( \frac{\sigma}{r}\right)^{6}\right],
\end{equation}
where $\varepsilon$ is the depth of the potential well and $\sigma$ is the distance at which the potential vanishes. The LJTS potential is
\begin{equation}
	\Phi_{\rm LJTS}(r)  = \left\{ \begin{array}{ll}
		\Phi_{\rm LJ}(r) - \Phi_{\rm LJ}(r_c) & r \le r_c \\
		0 & r > r_c
	\end{array} \right. .
\end{equation}
Although the two potentials differ only in the long-range attractive tail, the truncation has a significant effect on the thermodynamic properties, in particular on the location of the critical point; for this reason, the two systems are treated separately, each with its own EOS. We use LJ reduced units: $\rho^* = \rho\sigma^3$, $T^* = T/\varepsilon$ (with $k_B=1$), and $\lambda^* = \lambda\, \sigma^2\sqrt{m/\varepsilon}$ (in physical units, $\lambda^* = \lambda k_B^{-1} \sigma^2\sqrt{m/\varepsilon}$), where $m$ is the mass of a particle.

For both systems, the ideal thermal conductivity in Eq.\ \eqref{e.lambda_gen} is the dilute-gas value, obtained from the first-order Chapman-Enskog solution of the Boltzmann equation \cite{chapman,Hirschfelder},
\begin{equation}\label{e.lambdaid}
	\lambda_{\rm id}^* = \frac{75}{64} \sqrt{\frac{T^*}{\pi}}\frac{1}{\Omega^{(2,2)}},
\end{equation}
where $\Omega^{(2,2)}$ is the reduced collision integral, i.e., the collision integral normalized by its value for hard spheres of diameter $\sigma$. Higher-order Chapman-Enskog corrections for the LJ potential are below 1\% in the temperature range considered here \cite{Hirschfelder}, and are neglected. For both LJ and LJTS potentials we use the approximation of Fokin et al.\ \cite{fokin}:
 \begin{equation}\label{e.Omega22}
	\Omega^{(2,2)}(T^*) = \exp\left(-\frac{1}{6} \ln T^* + \ln{\frac{17}{18}} + \sum_{i=0}^{5} a_i {T^*}^{-i/2}\right),
\end{equation}
with $a_0=0.31081$, $a_1=-0.171211$, $a_2=-0.715805$, $a_3=2.48678$, $a_4=-1.78317$, $a_5=0.394405$. Equation \eqref{e.Omega22} was obtained for the full LJ potential; its use for the LJTS potential is an approximation, justified by the fact that binary collisions at the temperatures considered are dominated by the repulsive core and the region near the minimum of the potential, which are not affected by the truncation at $2.5\sigma$.

The thermodynamic quantities in Eqs.\ \eqref{e.lambda_gen} and \eqref{e.bracket}, namely $c_\rho$, $Z$, $\Gamma$, and their derivatives, are obtained from the Helmholtz free energy given by the EOS of each system. Since the theoretical result depends on the curvature of the entropy and on its variations, through derivatives of the specific heat, it is a stringent test not only of the transport theory but also of the EOS, whose higher-order derivatives are usually less accurate than the pressure or the internal energy.

We consider temperatures above the critical temperature of each system. The critical temperature of the LJ fluid is $T_c^* \simeq 1.32$ \cite{thol}, while for the LJTS fluid with $r_c=2.5\sigma$ it is significantly lower, $T_c^* \simeq 1.08$ \cite{vrabec2006}. The immediate vicinity of the critical point is avoided because the derivatives of the specific heat become very large there, so that the theoretical result is extremely sensitive to small variations of the curvature of the entropy; moreover, close to the critical point, long-range fluctuations invalidate the local-equilibrium picture of small, statistically independent cells on which the theory is based.

The theory applies only to a homogeneous, thermodynamically stable fluid: Eq.\ \eqref{e.L} relies on the Hessian of the thermodynamic potential, which must be negative definite, and on the local-equilibrium description of small cells of a single phase. The fluid EOS, on the other hand, does not contain the fluid-solid transition, and the changes of curvature of the free energy that appear when this transition is approached are not described by the theory. In order to show this limitation graphically, the figures of Sec.\ \ref{s.resLJ} and following that contain TTT results display in gray the region beyond the freezing line. For each temperature, the theoretical isotherm crosses the freezing line at the freezing density $\rho_{\rm f}^*(T^*)$, i.e., the density of the liquid in solid-liquid coexistence, where the theory gives the value $\lambda^*_{\rm f}(T^*) = \lambda^*_{\rm th}(\rho_{\rm f}^*(T^*),T^*)$. The parametric curve $(\rho_{\rm f}^*(T^*), \lambda_{\rm f}^*(T^*))$, starting at the triple point, assigns a freezing density to each value of $\lambda^*$, and the gray region is located on its high-density side: there, the theoretical isotherm that passes through each point has already crossed the freezing line. Below the triple point, the condensed phase is solid at any density, and the gray region is closed by a vertical line at $\rho_{\rm f}^*(T_{\rm t}^*)$, where $T_{\rm t}^*$ is the triple-point temperature. We used the solid-liquid coexistence data of Pedersen \cite{pedersen2013} for the LJ potential truncated at $r_c=6\sigma$, which is representative of the full LJ fluid \cite{mastny}, and for the LJTS potential with $r_c=2.5\sigma$, with triple-point temperatures $T_{\rm t}^* = 0.6946$ for LJ \cite{schultz2018} and $T_{\rm t}^*\simeq 0.62$ for LJTS. The freezing density increases with temperature, from $\rho_{\rm f}^*(T_{\rm t}^*) \simeq 0.845$ (LJ) and $0.83$ (LJTS) to values larger than 1 for $T^*\gtrsim 1.5$. The gray region is calculated with TTT and the same EOS used for the curves of each panel. For the temperatures considered in this work, $T^*\ge 1.35$, the freezing density is $\rho_{\rm f}^*\gtrsim 0.98$, and all the MD data used for the comparison, $\rho^*\le 0.8$, are in the homogeneous stable fluid, at a distance from the freezing line that increases with temperature. This is the reason why we restrict the comparison to densities $\rho^* \le 0.8$: although the EOS considered here are valid at larger densities, the theory is sensitive to the changes of curvature of the free energy that precede the phase transition, and the approach to the freezing line should be avoided.

To quantify the agreement between theory and simulations, we use the mean absolute percentage error (MAPE) for each isotherm,
\begin{equation}\label{e.MAPE}
	{\rm MAPE} = \frac{100\%}{n} \sum_{i=1}^{n} \left| \frac{\lambda_{{\rm th},i} - \lambda_{{\rm MD},i}}{\lambda_{{\rm MD},i}} \right|,
\end{equation}
where $\lambda_{{\rm MD},i}$ are the $n$ simulation values with $\rho^*\le 0.8$ along the isotherm and $\lambda_{{\rm th},i}$ are the corresponding theoretical predictions. All the figures and tables of this section were obtained with a SageMath notebook, which also contains the simulation data used for the comparison \cite{SM}.

\subsection{Choice of the equations of state}
\label{s.eos}

As mentioned above, Eq.\ \eqref{e.lambda_gen} requires derivatives of the Helmholtz free energy up to third order, and the result is therefore sensitive to the choice of the EOS. We have evaluated the theory with several EOS available in the literature: for the LJ fluid, the multiparameter EOS of Thol et al.\ \cite{thol}, the EOS of Kolafa and Nezbeda \cite{kolafa}, and the SAFT-VR Mie EOS of Lafitte et al.\ \cite{lafitte}; for the LJTS fluid, the multiparameter EOS of Thol et al.\ \cite{thol2} and the EOS of Heier et al.\ \cite{heier}. The smallest deviations from the simulation data were obtained with the SAFT-VR Mie EOS for the LJ fluid and with the EOS of Heier et al.\ for the LJTS fluid, and, in what follows, we use these two EOS. For the LJ fluid, we also show the results obtained with the multiparameter EOS of Thol et al.\ \cite{thol} for the data set of Fig.\ \ref{f.LJ_CM}, as an illustration of the sensitivity of the theory to the EOS.

Both EOS are based on Barker-Henderson perturbation theory \cite{barker} and share a feature that is relevant for the present application: they provide a physically consistent and detailed description of the metastable region, and of the continuation of the fluid branch inside the two-phase region. The EOS of Heier et al.\ was developed specifically to be used in theories of interfacial properties, such as density gradient theory, which require a reliable description of the fluid inside the metastable and unstable regions; it shows no unphysical behavior in these regions, while the EOS of Thol et al.\ for the LJTS fluid \cite{thol2}, which performs very well in the stable region, displays several van der Waals loops in the metastable and unstable regions \cite{heier}. The SAFT-VR Mie EOS is an analytic expression derived from perturbation theory, with a theoretically based dependence on density and temperature \cite{lafitte}. On the other hand, multiparameter EOS, such as those of Thol et al.\ \cite{thol}, are highly accurate for the properties to which they were fitted in the stable region, but the empirical functional forms used in their construction may produce oscillations in the higher-order derivatives of the free energy. The more detailed treatment of the metastable region apparently smooths the derivatives involved in Eq.\ \eqref{e.lambda_gen}, reducing the error of the prediction. For example, with the EOS of Thol et al.\ for the LJ fluid, the theory displays, at $T^*=1.36$, a pronounced spurious maximum at $\rho^*\simeq 0.3$, close to the critical density, that is not present in the simulation data [see Fig.\ \ref{f.LJ_CM}(b)], while such a feature is absent when the SAFT-VR Mie EOS is used.

An additional reason to choose the SAFT-VR Mie EOS for the LJ fluid is that it is the EOS used in the revised Enskog theories for Mie fluids of Jervell and Wilhelmsen \cite{jervell,jervell2024predicting}, which we use as a benchmark in Sec.\ \ref{s.resLJ}. In this way, both theoretical approaches are compared on the basis of the same thermodynamic input.

\subsection{Revised Enskog theory}
\label{s.RET}

In order to assess the performance of the thermodynamic transport theory (TTT) developed in this paper, Eq.\ \eqref{e.lambda_gen}, we compare its predictions with those of two versions of the revised Enskog theory (RET) for Mie fluids of Jervell and Wilhelmsen: the original one, RET-Mie \cite{jervell}, and the more recent RET-EWCA \cite{jervell2024predicting}. This kinetic theory extends the Enskog solutions of the Boltzmann equation, originally developed for hard spheres \cite{enskog,chapman}, to dense fluids of particles interacting through Mie potentials, of which the LJ potential is a particular case. In RET, the effect of the density on the transport coefficients enters through two mechanisms: the increase of the collision frequency, described by the radial distribution function (RDF) at contact, and the instantaneous transfer of momentum and energy across the collision diameter at the moment of the collision. In RET-Mie, the RDF at contact is obtained from the Barker-Henderson perturbation expansion of the SAFT-VR Mie EOS \cite{lafitte}, and the collision diameters are computed from the dynamics of binary collisions in the Mie potential \cite{jervell}. In RET-EWCA \cite{jervell2024predicting}, the collision diameter is replaced, in the residual (collisional) contributions, by transfer lengths for momentum and energy, which are obtained from the closest approach of colliding particles weighted by the momentum or energy exchanged in the collision, and averaged over the distribution of relative velocities; the attractive part of the potential is included in the scattering problem. The RDF at contact is the same as in RET-Mie. Neither version has adjustable parameters once the potential is specified.

Both TTT and RET give the thermal conductivity in terms of the dilute-gas value and equilibrium properties, but the underlying approaches are different: in RET, the equilibrium information enters through the RDF at contact, i.e., through structural properties, while in TTT it enters only through thermodynamic quantities obtained from the free energy. As a consequence, TTT can be used with any EOS, while RET requires an EOS, or another theory, that provides the RDF at contact. This is the reason why, for the LJTS fluid, we do not compare with RET: the EOS of Heier et al.\ does not provide the RDF required by the theory. Furthermore, RET-EWCA requires the solution of the scattering problem for each temperature to obtain the transfer lengths, while TTT only involves derivatives of the free energy.

The details of the calculation of the thermal conductivity in RET-Mie and RET-EWCA can be found in Refs.\ \cite{jervell,jervell2024predicting}. The only difference with respect to those procedures is that, instead of the explicit calculation of the collision integrals, we used the interpolation of Fokin et al., Eq.\ \eqref{e.Omega22}, which differs from the explicit calculation by about 0.1\%.

\subsection{Results for the LJ fluid}
\label{s.resLJ}

For the LJ fluid, we use two sets of molecular dynamics results. The first set was taken from Ref.\ \cite{chaparro}, where simulations of Mie fluids were carried out with a cutoff radius $r_c = 5\sigma$, which is considered large enough to represent the full LJ system; we consider the isotherms $T^* = 1.36$, $2$, $3.8$, $6$, and $7.2$. The second set corresponds to the results of Baidakov and Protsenko \cite{baidakov} for the LJ fluid with $r_c =6.78\sigma$, in stable and metastable states, for the isotherms $T^*=1.35$, $1.5$, and $2$. The lowest temperatures, $T^*=1.35$ and $1.36$, are close to the critical temperature, and the data of Ref.\ \cite{baidakov} include several points in the vicinity of the critical density; these isotherms therefore provide a demanding test of all the theories. Although the SAFT-VR Mie EOS does not specify explicit temperature limits for its validity, the interval $1.35 \le T^* \le 7.2$ provides a suitable domain for comparison with simulation data.

\begin{figure*}
	\includegraphics[width=0.329\textwidth]{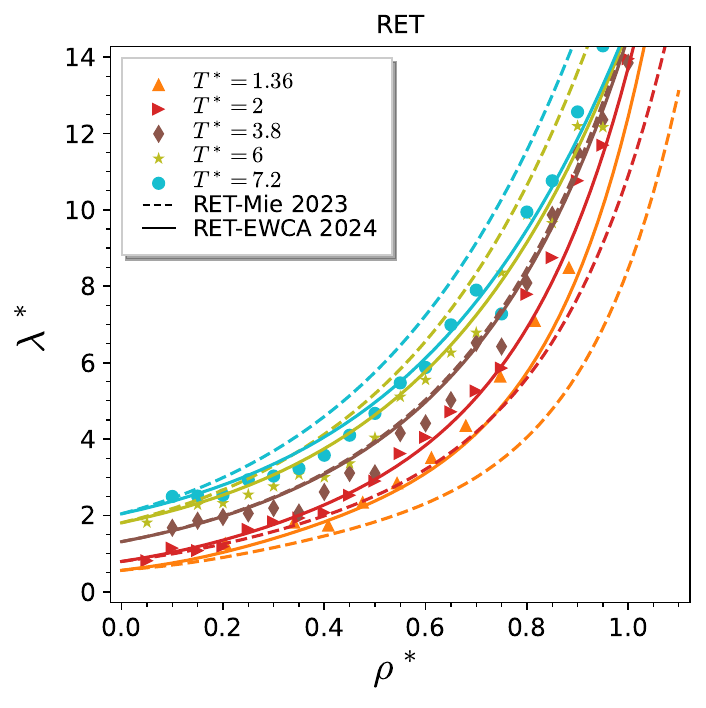}\hfill
	\includegraphics[width=0.329\textwidth]{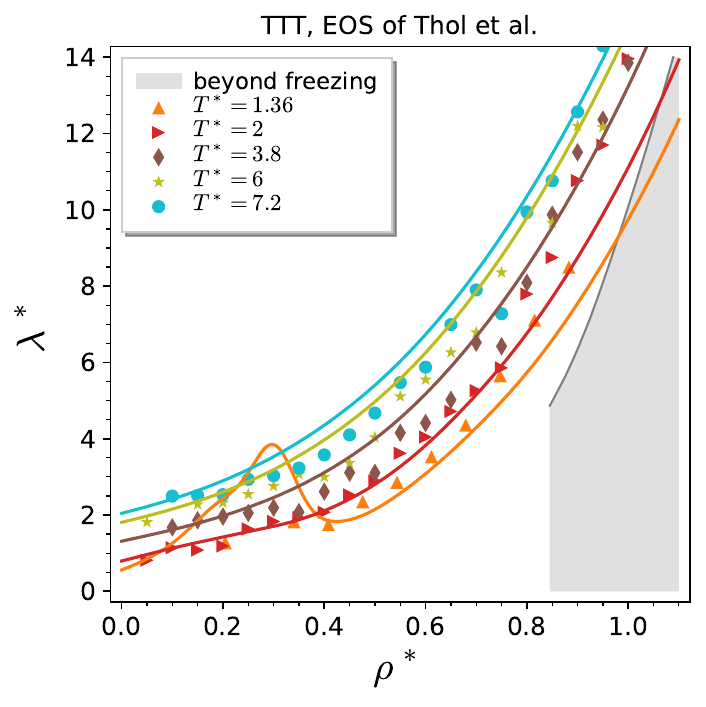}\hfill
	\includegraphics[width=0.329\textwidth]{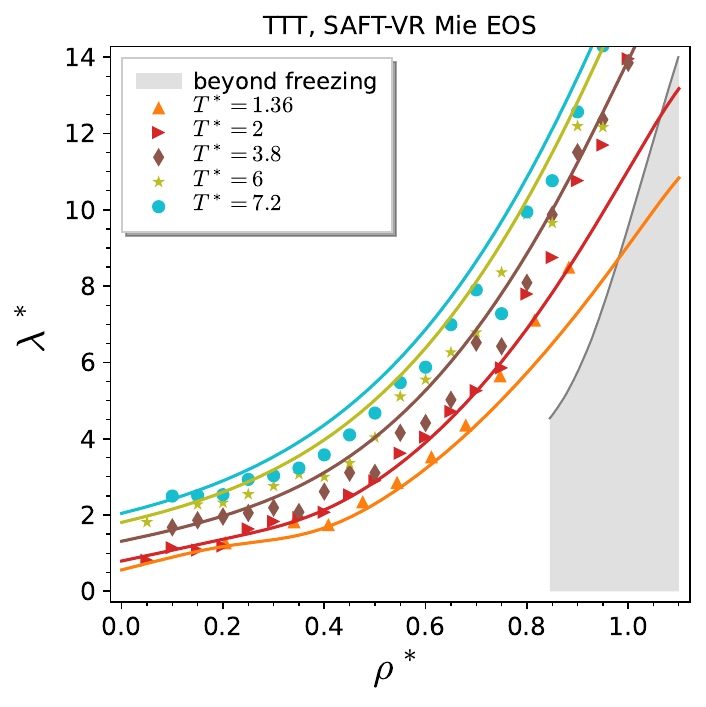}
	\caption{Reduced thermal conductivity $\lambda^*$ of the LJ fluid as a function of density for the temperatures indicated in the legends. Symbols: molecular dynamics results of Chaparro and M\"uller \cite{chaparro}. (a, left panel) Dashed lines: RET-Mie \cite{jervell}; solid lines: RET-EWCA \cite{jervell2024predicting}; in both cases, the RDF at contact is given by the SAFT-VR Mie EOS \cite{lafitte}. (b, middle panel) Solid lines: TTT, Eq.\ \eqref{e.lambda_gen}, with the EOS of Thol et al.\ \cite{thol}. (c, right panel) Solid lines: TTT with the SAFT-VR Mie EOS \cite{lafitte}. In TTT, $\lambda_{\rm id}$ is given by Eq.\ \eqref{e.lambdaid}. In panels (b) and (c), the gray regions lie beyond the freezing line (see Sec.\ \ref{s.comp}).}
	\label{f.LJ_CM}
\end{figure*}

\begin{table}
	\caption{MAPE (in \%), Eq.\ \eqref{e.MAPE}, of RET-Mie, RET-EWCA, and TTT with respect to the simulation results of Chaparro and M\"uller \cite{chaparro} for the LJ fluid, for $\rho^* \le 0.8$ (see Fig.\ \ref{f.LJ_CM}). TTT Thol and TTT Lafitte correspond to the EOS of Thol et al.\ \cite{thol} and to the SAFT-VR Mie EOS of Lafitte et al.\ \cite{lafitte}, respectively.}
	\label{t.LJ_CM}
	\begin{ruledtabular}
	\begin{tabular}{ccccc}
		$T^*$ & RET-Mie & RET-EWCA & TTT Thol & TTT Lafitte \\
		\hline
		1.36 & 29 & 11 & 28 & 8.8 \\
		2 & 14 & 6.6 & 7.5 & 6.6 \\
		3.8 & 12 & 11 & 13 & 14 \\
		6 & 19 & 9.2 & 14 & 15 \\
		7.2 & 20 & 7.1 & 13 & 14 \\
	\end{tabular}
	\end{ruledtabular}
\end{table}

The comparison with the data of Chaparro and M\"uller is shown in Fig.\ \ref{f.LJ_CM}, and the corresponding values of the MAPE are given in Table \ref{t.LJ_CM}. In the low-density limit, all the theories tend to the dilute-gas value, in agreement with the simulation data, which confirms the adequacy of Eqs.\ \eqref{e.lambdaid} and \eqref{e.Omega22}. At higher densities, RET-Mie predicts a monotonic increase of $\lambda^*$ with density, with positive curvature, but systematically underestimates the thermal conductivity at low temperatures and overestimates it at high temperatures; its MAPE is between 12\% and 29\%. The energy-transfer lengths of RET-EWCA correct most of these deviations: its MAPE is about 11\% at $T^*=1.36$ and 3.8, and between 6.6\% and 9.2\% at the other temperatures, i.e., two to three times smaller than that of RET-Mie, except at $T^*=3.8$, where both versions give similar errors. TTT with the SAFT-VR Mie EOS captures the weak density dependence at low and intermediate densities for the lowest temperatures, where an inflection appears in the isotherms, as well as the steep increase of $\lambda^*$ in the dense fluid, where the thermal conductivity reaches values about ten times larger than $\lambda_{\rm id}^*$. At $T^* = 1.36$ and 2, its MAPE, 8.8\% and 6.6\%, is similar to, or slightly smaller than, that of RET-EWCA. At higher temperatures, TTT tends to overestimate the thermal conductivity at intermediate and high densities, with MAPE values of 14--15\%, larger than those of RET-EWCA but smaller than those of RET-Mie at $T^*=6$ and 7.2. With the EOS of Thol et al., the results of TTT are similar to those obtained with the SAFT-VR Mie EOS at $T^*\ge 2$, but, at $T^*=1.36$, the spurious maximum discussed in Sec.\ \ref{s.eos} raises the MAPE to 28\%. At low densities and high temperatures, the scatter of the simulation data is comparable to the differences between theory and simulation. In all cases, the TTT curves cross the boundary of the gray region at densities well above $\rho^*=0.8$.

\begin{figure*}
	\includegraphics[width=0.49\textwidth]{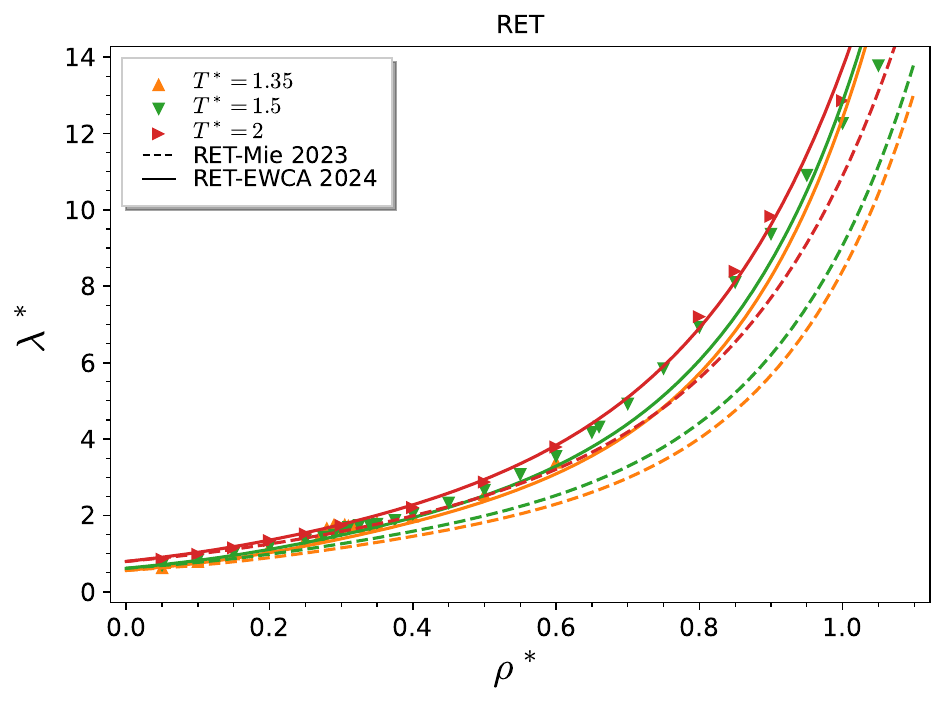}\hfill
	\includegraphics[width=0.49\textwidth]{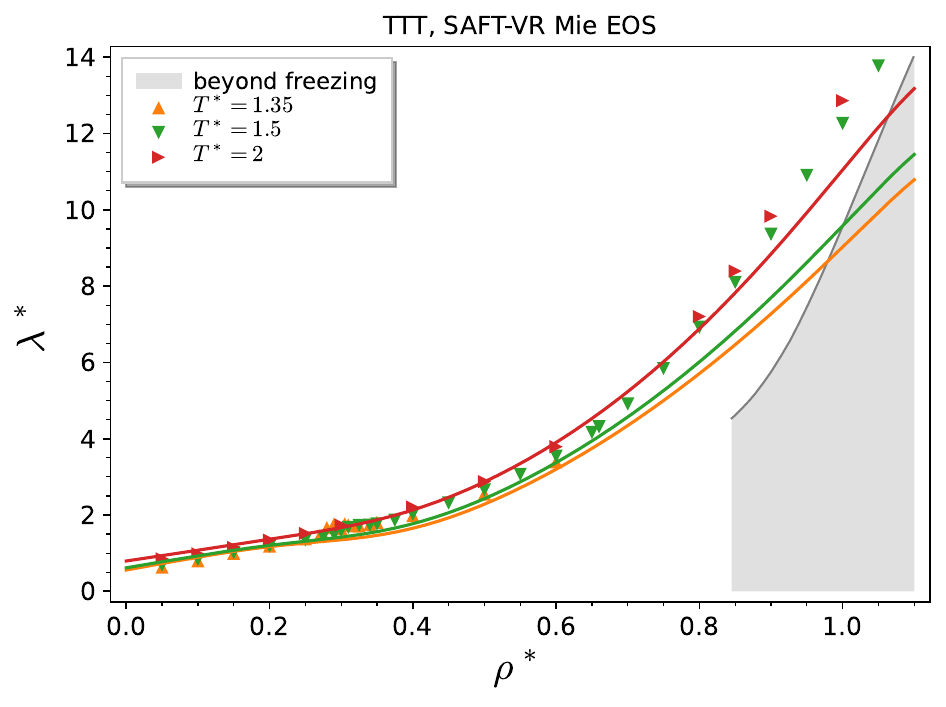}
	\caption{Same as Fig.\ \ref{f.LJ_CM}, but with the molecular dynamics results of Baidakov and Protsenko \cite{baidakov} (symbols). (a, left panel) Dashed lines: RET-Mie \cite{jervell}; solid lines: RET-EWCA \cite{jervell2024predicting}. (b, right panel) Solid lines: TTT, Eq.\ \eqref{e.lambda_gen}. Thermodynamic input given by the SAFT-VR Mie EOS \cite{lafitte}. In panel (b), the gray region lies beyond the freezing line.}
	\label{f.LJ_BP}
\end{figure*}

\begin{table}
	\caption{MAPE (in \%), Eq.\ \eqref{e.MAPE}, of RET-Mie, RET-EWCA, and TTT with respect to the simulation results of Baidakov and Protsenko \cite{baidakov} for the LJ fluid, for $\rho^* \le 0.8$ (see Fig.\ \ref{f.LJ_BP}). In all cases, the thermodynamic input is given by the SAFT-VR Mie EOS.}
	\label{t.LJ_BP}
	\begin{ruledtabular}
	\begin{tabular}{cccc}
		$T^*$ & RET-Mie & RET-EWCA & TTT \\
		\hline
		1.35 & 27 & 13 & 15 \\
		1.5 & 22 & 6.2 & 8.8 \\
		2 & 9.1 & 2.9 & 4.2 \\
	\end{tabular}
	\end{ruledtabular}
\end{table}

The comparison with the data of Baidakov and Protsenko is shown in Fig.\ \ref{f.LJ_BP} and Table \ref{t.LJ_BP}. The errors obtained for RET-Mie in Table \ref{t.LJ_BP} are consistent with those reported by Jervell and Wilhelmsen \cite{jervell} (see Fig.\ S.13 in their Supporting Information). These results confirm the trends observed with the previous data set. RET-Mie underestimates the thermal conductivity at intermediate and high densities, with MAPE values of 27\% and 22\% at $T^* = 1.35$ and 1.5, respectively. RET-EWCA and TTT reproduce the shape of the isotherms, and their errors are similar: the MAPE of RET-EWCA is 13\%, 6.2\%, and 2.9\%, and that of TTT is 15\%, 8.8\%, and 4.2\%, at $T^*=1.35$, 1.5, and 2, respectively, i.e., about one half, or less, of the error of RET-Mie. The largest error of TTT, 15\% at $T^*=1.35$, is found at the temperature closest to the critical point, where the theory slightly underestimates the simulation data at intermediate densities, $0.4 \lesssim \rho^* \lesssim 0.6$; RET-EWCA has a similar error at this temperature. As discussed above, close to the critical point the result is very sensitive to the curvature of the entropy, and the local-equilibrium assumptions of the theory become less accurate; moreover, the simulation data may contain contributions of the critical enhancement of the thermal conductivity, which are not described by any of the theories considered here.

\subsection{Results for the LJTS fluid}

For the LJTS fluid we use the EOS of Heier et al.\ \cite{heier}, and two sets of molecular dynamics results: our own simulations, for $T^*=1.35$, 1.5, 2, 4, and 6, and densities between $\rho^*=0.05$ and 0.8, and the results of Galliero and Boned \cite{galliero}, for $T^* = 1.5$, 2, 3, 4, and 6, and densities between $\rho^*=0.1$ and 0.8. As explained in Sec.\ \ref{s.RET}, a comparison with RET is not possible in this case.

Our molecular dynamics simulations were performed using LAMMPS (Large-scale Atomic/Molecular Massively Parallel Simulator)\cite{lammps}. We employed 500,000 particles initially arranged on an fcc lattice, with a lattice spacing chosen to achieve the desired density. Periodic boundary conditions were applied and time integration was performed using the Verlet algorithm implemented in LAMMPS, with a time step of $\Delta t^* = 0.001$. The system was equilibrated in two stages. First, particle velocities were assigned from a Gaussian distribution, and the system was evolved for 400,000 time steps in the NVT ensemble. The ensemble was then switched to NVE, and the system was evolved for a further 400,000 time steps. Following equilibration, production simulations were performed for $4\text{--}10 \times 10^6$ time steps in the NVE ensemble. The thermal conductivity was calculated using the Green--Kubo method, which relates it to the time integral of the autocorrelation function of the heat flux (see Ch. 21 in \cite{mcquarrie}). The numerical values are given in Ref.\ \cite{SM}.

\begin{figure}
	\includegraphics[width=\columnwidth]{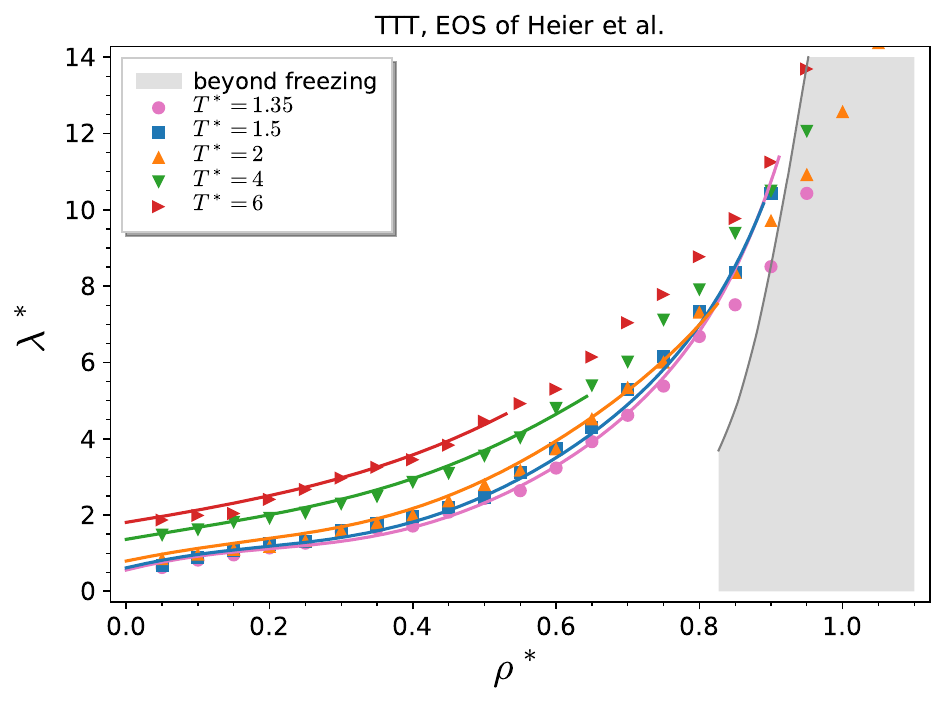}
	\caption{Reduced thermal conductivity $\lambda^*$ of the LJTS fluid ($r_c=2.5\sigma$) as a function of density for the temperatures indicated in the legend. Symbols: our molecular dynamics results. Lines: TTT, Eq.\ \eqref{e.lambda_gen}, with thermodynamic quantities obtained from the EOS of Heier et al.\ \cite{heier}. The gray region lies beyond the freezing line.}
	\label{f.LJTS}
\end{figure}

\begin{table}
	\caption{MAPE (in \%), Eq.\ \eqref{e.MAPE}, of TTT with respect to our simulation results for the LJTS fluid (see Fig.\ \ref{f.LJTS}).}
	\label{t.LJTS}
	\begin{ruledtabular}
	\begin{tabular}{cc}
		$T^*$ & TTT \\
		\hline
		1.35 & 6.4 \\
		1.5 & 6.6 \\
		2 & 7.7 \\
		4 & 5.2 \\
		6 & 4.0 \\
	\end{tabular}
	\end{ruledtabular}
\end{table}

\begin{figure}
	\includegraphics[width=\columnwidth]{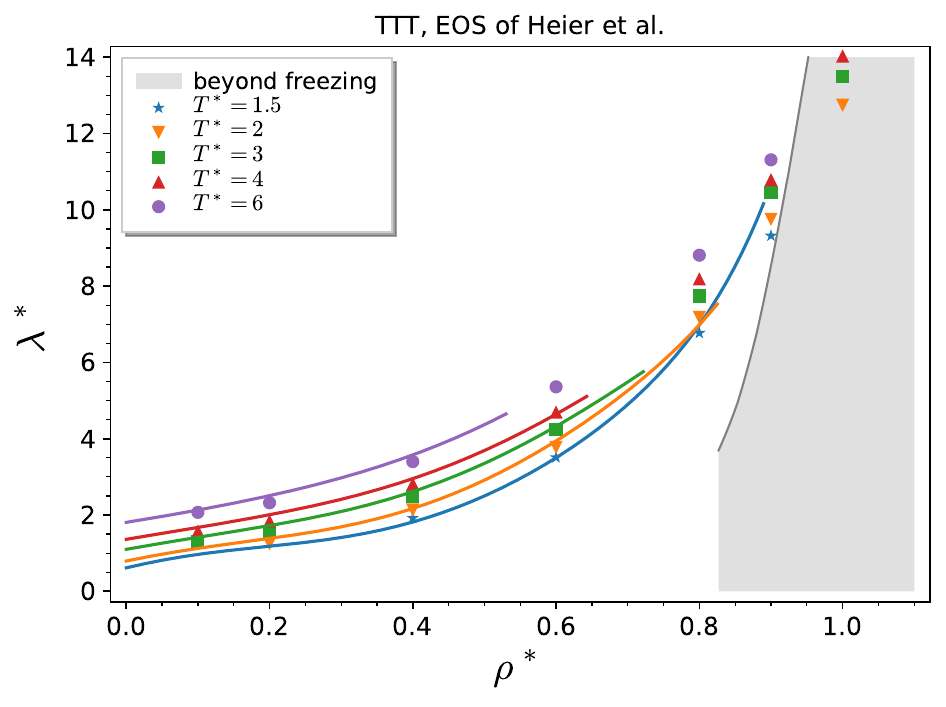}
	\caption{Same as Fig.\ \ref{f.LJTS}, but with the molecular dynamics results of Galliero and Boned \cite{galliero} (symbols).}
	\label{f.LJTS_G}
\end{figure}

\begin{table}
	\caption{MAPE (in \%), Eq.\ \eqref{e.MAPE}, of TTT with respect to the simulation results of Galliero and Boned \cite{galliero} for the LJTS fluid (see Fig.\ \ref{f.LJTS_G}).}
	\label{t.LJTS_G}
	\begin{ruledtabular}
	\begin{tabular}{cc}
		$T^*$ & TTT \\
		\hline
		1.5 & 2.9 \\
		2 & 5.0 \\
		3 & 7.7 \\
		4 & 7.0 \\
		6 & 4.9 \\
	\end{tabular}
	\end{ruledtabular}
\end{table}

The comparison with our simulation results is shown in Fig.\ \ref{f.LJTS} and Table \ref{t.LJTS}, and the comparison with the data of Galliero and Boned in Fig.\ \ref{f.LJTS_G} and Table \ref{t.LJTS_G}. Both sets of numerical results are consistent with each other at the common temperatures. At high temperatures, the theoretical curves are shown only up to a maximum density that decreases as the temperature increases. This limit corresponds to the restriction of the application of the EOS of Heier et al.\ to reduced pressures $p^* < 6.8$, where $p^* = p\sigma^3/\varepsilon$. This conservative restriction follows the range of validity of the EOS of Thol et al.\ \cite{thol2}, $0.64 < T^* < 11$ and $p^* < 6.8$, which is determined by the simulation data used in its construction. Heier et al.\ validated their EOS in the stable region by comparison with these simulation data, as well as with the EOS of Thol et al.\ \cite{heier}. The restriction therefore keeps our calculations within the range in which the EOS of Heier et al.\ was tested in the homogeneous stable region; it should not be interpreted as an intrinsic upper limit of the pressure range of this EOS.

The agreement between TTT and the simulation results is very good for the LJTS fluid. The MAPE is between 4.0\% and 7.7\% for our data, and between 2.9\% and 7.7\% for the data of Galliero and Boned, with no systematic dependence on temperature. As in the LJ case, the low-density limit is correctly reproduced, and the inflection of the isotherms at intermediate densities, present for the lowest temperatures, is also captured by the theory, as well as the steep increase of the thermal conductivity in the dense fluid. The deviations do not show a systematic sign along the isotherms, and they are comparable for the two independent sets of simulation data. It is worth noting that the errors for the LJTS fluid at $T^* = 1.35$ and $1.5$, whose reduced distance to the critical point, $T/T_c$, is larger than for the LJ fluid at the same temperatures, are significantly smaller than those found for the LJ fluid. This is consistent with the observation that the largest errors of the theory are associated with the proximity of the critical point.

Overall, considering that Eq.\ \eqref{e.lambda_gen} contains no adjustable parameters, and that the thermal conductivity varies by about one order of magnitude in the range of densities considered, the agreement between theory and simulations is remarkable for both the LJ and the LJTS systems. For the LJ fluid, the accuracy of TTT is similar to that of RET-EWCA, the most accurate version of the kinetic theory considered here, and much better than that of RET-Mie.

\section{Conclusions}
\label{s.conclusions}

We have obtained a general expression for the thermal conductivity of a simple fluid, Eq.\ \eqref{e.lambda_gen}, using the thermodynamic transport theory of Ref.\ \cite{dimuro-hoyuelos}, in which the Onsager coefficients of an interacting system are related to those of an ideal reference system through the ratio of the determinants of the Hessians of the relevant thermodynamic potential. The main ingredient of the derivation is the identification of the transported quantity: instead of the full internal energy density, we considered an effective energy density proportional to the rms energy fluctuations of a small cell, whose dependence on the thermodynamic state differs from that of the internal energy. Since the particle density is not transported diffusively, it was eliminated through a Legendre transform, so that the problem reduces to a single transported variable.

The result expresses the thermal conductivity as the product of its dilute-gas value, $\lambda_{\rm id}$, and a function of equilibrium thermodynamic properties, namely the specific heats at constant $\rho$ and at constant $\rho T$ and the temperature derivative of the latter. The expression contains no adjustable parameters and reduces to the previously verified result for hard spheres, $\lambda/\lambda_{\rm id} = (2\Gamma + 3)/5$, when the specific heat is that of the ideal gas and the compressibility factor is independent of temperature.

The theory, which we called TTT, was tested against molecular dynamics results for the LJ and LJTS fluids in the supercritical region and for densities up to $\rho^* = 0.8$. The theory applies only to a homogeneous stable fluid; the region beyond the freezing line, shown in gray in the figures, illustrates this limitation and justifies the upper limit of the density range. Since the result involves derivatives of the free energy up to third order, it is sensitive to the EOS; the best results were obtained with EOS based on perturbation theory, the SAFT-VR Mie EOS for the LJ fluid and the EOS of Heier et al.\ for the LJTS fluid, whose detailed treatment of the metastable region apparently yields smoother derivatives than multiparameter EOS. With these EOS, TTT reproduces the shape of the isotherms, including the weak density dependence at low and intermediate densities and the steep increase in the dense fluid. For the LJTS fluid, the MAPE is below 8\% for all the temperatures considered. For the LJ fluid, TTT was compared with two versions of the revised Enskog theory for Mie fluids, RET-Mie and RET-EWCA, using the same EOS. The errors of TTT are similar to those of RET-EWCA: slightly smaller or equal at the lowest temperatures of the data set of Chaparro and M\"uller, slightly larger for the data of Baidakov and Protsenko, and larger at the highest temperatures, where the MAPE of TTT is about 14\%. With respect to RET-Mie, the errors of TTT are smaller at all temperatures except $T^* = 3.8$, and are reduced by a factor of two to three at low temperatures. The largest errors of TTT are found close to the critical point, where the result becomes very sensitive to the curvature of the entropy.

It is remarkable that TTT and RET-EWCA give errors of similar magnitude following very different theoretical routes. RET-EWCA is a kinetic theory: it requires the RDF at contact and the solution of the two-body scattering problem to obtain transfer lengths for momentum and energy that depend on the details of the collisions in the attractive and repulsive parts of the potential. In TTT, the dynamics enters only through the dilute-gas conductivity, and the effect of the density is given by a function of equilibrium thermodynamic properties, without any information on the structure of the fluid or on the collision dynamics beyond the dilute limit.

A practical advantage of TTT is that it only requires thermodynamic information, which can be obtained from any EOS, while kinetic approaches such as RET also require structural information, such as the RDF at contact, and, in the case of RET-EWCA, the numerical solution of the scattering problem at each temperature. This is also the case of the recently proposed REST framework \cite{jervell2026}, which, like TTT, aims at predicting transport properties without fitting parameters to transport data. In REST, the thermal conductivity is computed with RET-Mie in a region of high temperatures and moderate residual entropies, where RET is considered reliable, and is then extrapolated to the liquid-like region through a residual-entropy scaling relation, whose coefficients are fixed by continuity conditions at the boundary of that region. This procedure involves some additional choices: the boundary of the region where RET is reliable is selected by inspection of the RET predictions for model fluids, the functional form of the extrapolation is assumed, and the scaling incorporates a crossover function taken from previous empirical work. In TTT, instead, the thermal conductivity follows directly from Eq.\ \eqref{e.lambda_gen} at each thermodynamic state, without cutoffs or extrapolation schemes.

These results support the idea that a substantial part of the effect of the interactions on transport coefficients can be captured by equilibrium thermodynamics, with the dynamics contained in the ideal reference coefficient: for the thermal conductivity of simple fluids, this purely thermodynamic description reaches an accuracy comparable to that of the most recent version of the revised Enskog theory. Possible extensions of this work include the study of the near-critical region, where a description of the critical enhancement of the thermal conductivity would require including the effect of long-range fluctuations; and the application of the method to mixtures, where the coupling between heat and mass transport must be considered.

\begin{acknowledgments}
This work was partially supported by the Universidad Nacional de Mar del Plata (UNMdP, Argentina, Grant No. 80020250500008MP). Simulations were carried out using the HPC-UNMDP computing cluster. The AI tool Claude Opus 5.5 (Anthropic) was used to draft and revise parts of the text and of the SageMath notebook \cite{SM}; the authors reviewed and take full responsibility for the content.
\end{acknowledgments}

\appendix

\section{Hessian matrix}
\label{app.hessian}

We give here the explicit Hessian matrix used in Sec.~\ref{s.calc}. For simplicity, we write the differential of $\phi$, Eq.\ \eqref{e.dphi}, as
\begin{equation}
	d\phi=\tilde{Y}\,d\tilde{x} + \rho\,d\mu_T',
\end{equation}
where
\begin{equation}
	\mu_T' = \mu_T + c_\rho,	\qquad
	\tilde{Y}=\frac{c_\rho}{\epsilon T}.
\end{equation}
The Hessian of $\phi$ with respect to its natural variables is
\begin{equation}
	H_\phi = \frac{\partial^2 \phi}{\partial (\tilde{x},\mu_T')^2} = \frac
	{\partial(\tilde{Y},\rho)}{\partial(\tilde{x},\mu_T')},
\end{equation}
where the right-hand side denotes the Jacobian matrix of the transformation from $(\tilde{x},\mu_T')$ to $(\tilde{Y},\rho)$. Since $\tilde{Y}$ and $\rho$ are explicit functions of $\rho$ and $T$, it is simpler to calculate its inverse,
\begin{equation}
	H_\phi^{-1}
	=
	\frac{\partial(\tilde{x},\mu_T')}
	{\partial(\tilde{Y},\rho)},
\end{equation}
by writing all quantities in terms of $\rho$ and $T$ and using the chain rule. With $\tilde{x}=\epsilon T\rho$, the result is
\begin{align}
	& H_\phi^{-1} =  \left(c_\rho - T\cvT\right)^{-1}\times\nonumber\\
	&\begin{pmatrix}
		-\epsilon^2 T^2 \rho
		&
		\epsilon T\left( \cvrr - T \cvT\right)
		\\[7pt]
		\epsilon T\left( \cvrr - T \cvT \right)
		&
		\left(\frac{c_\rho}{\rho} - \frac{T}{\rho}\cvT \right)\difGT - \left(\cvr\right)^2 \rho
	\end{pmatrix},
	\label{e.Hinv}
\end{align}
where we have also used the following thermodynamic relations:
\begin{equation}\label{e.mutder}
	\left(\frac{\partial\mu_T}{\partial T}\right)_\rho
	=-\frac{1}{T}\cvrr,
	\qquad
	\left(\frac{\partial\mu_T}{\partial\rho}\right)_T
	=\frac{1}{\rho} \difGT.
\end{equation}
The first relation is obtained from the equality of the mixed second derivatives of the entropy density, using $(\partial s/\partial T)_\rho = \rho c_\rho/T$ and $(\partial s/\partial \rho)_T = -\mu_T$ [see Eq.\ \eqref{e.durT}]. The second one follows from the definition of the thermodynamic factor, $\Gamma = \rho\, \partial(\mu/T)/\partial\rho = \partial (\rho Z)/\partial \rho$, which gives $\partial\mu/\partial\rho = T\Gamma/\rho$; the equality between both expressions of $\Gamma$ is a consequence of the Gibbs-Duhem relation at constant temperature, $dP = \rho\, d\mu$.

The determinant of the Hessian is
\begin{equation}\label{e.detH}
	\det H_\phi
	= -\frac{c_\rho - T\cvT}{\epsilon^2 T^2 \left(2 \rho \cvr + \difGT + c_\rho - T \cvT\right)}.
\end{equation}
This expression can be simplified using the specific heat at constant $\rho T$, defined in terms of the entropy per particle, $s'$, as
\begin{align}
	c_{\rho T} &= T \left(\frac{\partial s'}{\partial T}\right)_{\rho T} = T \left(\frac{\partial s'}{\partial T}\right)_{\rho} - \rho \left(\frac{\partial s'}{\partial \rho}\right)_T \nonumber \\
	&= c_\rho + \frac{1}{\rho} \frac{\partial P}{\partial T} = c_\rho +  \frac{\partial (ZT)}{\partial T},
\end{align}
where the operator identity \eqref{e.operator} and the Maxwell relation $(\partial s'/\partial \rho)_T = -\rho^{-2} (\partial P/\partial T)_\rho$ were used. Applying the operator \eqref{e.operator} to $c_{\rho T}$, we obtain
\begin{equation}
	-T \left(\frac{\partial c_{\rho T}}{\partial T}\right)_{\rho T} = 2 \rho \cvr  + \difGT + c_\rho - T \cvT - c_{\rho T},
\end{equation}
which is equivalent to Eq.\ \eqref{e.bracket}. In this derivation we have used that $\partial (\Gamma T)/\partial T = \partial (ZT)/\partial T + \rho\, \partial^2 (ZT)/\partial \rho\, \partial T$ and the relation $-T\, \partial^2 (ZT)/\partial T^2 = \rho\, \partial c_\rho/\partial \rho$, which is the density version of the well-known identity $(\partial C_V/\partial V)_T = T (\partial^2 P/\partial T^2)_V$. Equation \eqref{e.detH} then becomes
\begin{equation}\label{e.detH2}
	\det H_\phi
	= -\frac{c_\rho - T\cvT}{\epsilon^2 T^2 \left[c_{\rho T} - T \left(\frac{\partial c_{\rho T}}{\partial T}\right)_{\rho T}\right]}.
\end{equation}

In the ideal case, $c_\rho=c_\rho^{\rm id}$ is constant and $c_{\rho T}^{\rm id}=c_\rho^{\rm id} + 1$ (since $Z=1$), so that
\begin{equation}
	\det H_\phi^{\rm id}
	= -\frac{c_\rho^{\rm id}}{\epsilon^2 T^2 (c_\rho^{\rm id} + 1)}.
\end{equation}
Therefore,
\begin{equation}
	\frac{\det H_\phi^{\rm id}}
	{\det H_\phi}
	= \frac{c_\rho^{\rm id} \left[c_{\rho T} - T \left(\frac{\partial c_{\rho T}}{\partial T}\right)_{\rho T}\right]}{(c_\rho^{\rm id} + 1)\left(c_\rho - T\cvT\right)},
\end{equation}
which leads directly to Eq.~\eqref{e.LLid} through Eq.~\eqref{e.L}.

\bibliography{LJth_cond.bib}

\end{document}